\documentclass[11pt]{article}
\usepackage[
]{graphicx}
\usepackage{amsmath,amssymb}
\usepackage{cases}
\usepackage{color}
\usepackage[
]{hyperref}

\baselineskip=\normalbaselineskip
\renewcommand{\baselinestretch}{1.4}
\newcommand{\resection}[1]
 {\setcounter{equation}{0}\section{\large{#1}}}

\begin{document}


\setcounter{page}{0}
\begin{flushright}
July 2019\\
RUP-19-2 
\end{flushright}

\vfill

\begin{center}
{\Large{\bf 
Null Geodesics and Repulsive Behavior of Gravity \\
in $(2+1)$-dimensional Massive Gravity 
}}
\end{center}

\vfill

\renewcommand{\baselinestretch}{1.0}

\begin{center}

{\sc Keisuke Nakashi}
\footnote{E-mail:{\tt nakashi@rikkyo.ac.jp}}, 
{\sc Shinpei Kobayashi}
\footnote{E-mail:{\tt shimpei@u-gakugei.ac.jp}},
{\sc Shu Ueda}
\footnote{E-mail:{\tt m171805s@st.u-gakugei.ac.jp}} 
and \ 
{\sc Hiromi Saida}
\footnote{E-mail:{\tt saida@daido-it.ac.jp}} 

~\\
$^{1}${\sl Department of Physics, Rikkyo University, \\ 
    Toshima, Tokyo 171-8501, JAPAN }  \\
     \vspace{1cm}

$^{2, 3}${\sl Department of Physics, Tokyo Gakugei University, \\ 
     4-1-1 Nukuikitamachi, Koganei, Tokyo 184-8501, JAPAN }  \\
     \vspace{1cm}
$^{4}${\sl Department of Physics, Daido University, \\ 
     Minami-ku, Nagoya, 457-8530, JAPAN } 

\end{center}

\vfill

\begin{center}
{\bf abstract}
\end{center}

\begin{quote}

\small{%
\ \ \ We study the null geodesics in a static circularly symmetric (SCS) black hole spacetime which is a solution in the $(2+1)$-dimensional massive gravity proposed by Bergshoeff, Hohm and Townsend (BHT massive gravity). 
We obtain analytic solutions for the null geodesic equation in the SCS black hole background and find the explicit form of deflection angles. 
We see that for various values of the impact parameter, the deflection angle can be positive, negative or even zero in this black hole spacetime. 
The negative deflection angle indicates the repulsive behavior of the gravity which comes from the gravitational hair parameter that is the most characteristic quantity of the BHT massive gravity.
}
\end{quote}
\vfill

\renewcommand{\baselinestretch}{1.4}

\renewcommand{\thefootnote}{\arabic{footnote}}
\setcounter{footnote}{0}
\addtocounter{page}{1}
\newpage

\resection{Introduction}
\ \ \ It is well known that in $(2+1)$ dimensions, the Riemann tensor can be expressed in terms of the Ricci tensor and the Ricci scalar since the Weyl tensor is identically zero. 
This fact means that there are no local degrees of freedom. 
Due to this simplicity, Einstein gravity in $(2+1)$ dimensions does not have a nontrivial black hole solution except for the BTZ black hole solution \cite{Banados:1992wn, Banados:1992gq}, which is the solution of the Einstein equation with a negative cosmological constant. 

In addition, there are some modifications to introduce local degrees of freedom to gravity in $(2+1)$ dimensions. 
For example, the topologically massive gravity \cite{Deser:1981wh}, which has the Lorentz Chern--Simons term in its action in addition to the Einstein--Hilbert action, possesses a single propagating degree of freedom in the linearization level around maximally symmetric spacetime. 

Another massive gravity in $(2+1)$ dimensions has been suggested by  Bergshoeff, Hohm and Townsend (BHT) \cite{Bergshoeff:2009hq, Bergshoeff:2009aq}. 
The BHT massive gravity is a ghost-free theory with quadratic terms of the Ricci tensor and Ricci scalars by adjusting the coefficients appropriately \cite{Deser:2009hb}. 
The BHT massive gravity has some nontrivial black hole solutions \cite{Oliva:2009ip}, including the BTZ black hole as its special case. 

Since the BHT massive gravity includes a massive graviton which gives a new mass scale, we can expect that the large-scale interaction of gravity must be different from that of Einstein gravity. 
The deviation from Einstein gravity appears as a new parameter in the black hole solution, which is called the gravitational hair parameter \cite{Oliva:2009ip}. 
In order to clarify the consequences of the deviation, we investigate the null geodesics and the deflection angles of the null geodesics in the black hole spacetime with the gravitational hair parameter. 
\footnote{For the $(2+1)$-dimensional black hole in Einstein gravity, i.e., the BTZ black hole, the analytic solution for the geodesic equation for massless particles are examined in \cite{Cruz:1994ir} and for $(3+1)$ and higher-dimensional black holes, these are discussed in \cite{Hackmann:2008zz,Hackmann:2010tqa}.}
We obtain analytic solutions for the geodesic equation of massless particles and find that the gravity behaves as if a repulsive force acts on the geodesics with some values of the parameters. 
Instead of the effective potential, we calculate the deflection angles of the null geodesics to evaluate this repulsive behavior of the gravity and then we obtain the explicit form of the deflection angles in the black hole background. 
We show that the deflection angle in the SCS black hole spacetime can be negative. 
This fact indicates the repulsive behavior of the gravity. 
The negative deflection angles caused by wormholes are mentioned in \cite{Cramer:1994qj,Shaikh:2016dpl,Shaikh:2017zfl}. 
The applications of such behaviors of null geodesics around exotic objects to the gravitational lensing have also already been discussed in~\cite{Kitamura:2012zy,Izumi:2013tya}.
Also, we show that the origin of the repulsive behavior of the BHT massive gravity is the existence of the gravitational hair parameter. 
In fact, it is known that the Lifshitz black hole \cite{AyonBeato:2009nh,Cruz:2013ufa}, which is another type of black hole spacetime in the BHT massive gravity, does not have the gravitational hair parameter and the geodesics in its black hole background do not show such a repulsive behavior of the gravity. 
This is consistent with our claim. 

This paper is organized as follows. In Section 2, we explain the BHT massive gravity and the SCS black hole spacetime. 
In Section 3, we derive the analytic solution of the geodesic equation for massless particles in the SCS black hole spacetime. 
Also, we discuss that there exist geodesics whose behaviors seem as if the massless particles receive a repulsive force from the black hole. 
In Section 4, we introduce the deflection angles of the null geodesics in the SCS black hole spacetime. 
We derive the explicit form of the deflection angle and reveal that, for various values of the impact parameter, 
the deflection angles can be positive, negative, or even zero. 
The last section is devoted to the conclusion and the discussion.  

\resection{BHT massive gravity and static circularly symmetric black hole solution}
\ \ \ The BHT massive gravity is characterized by the following action \cite{Bergshoeff:2009hq}
\begin{eqnarray}
S = \frac{1}{16 \pi G}\int d^{3}x \sqrt{-g} \left (R - 2\lambda - \frac{1}{m^2}K \right ),
\label{BHTaction}
\end{eqnarray}
where $K$ is quadratic in the Ricci tensor and the Ricci scalar as 
\begin{eqnarray}
K = R_{\mu \nu }R^{\mu \nu }-\frac{3}{8}R^{2}.
\end{eqnarray}
The source-free field equation can be obtained as
\begin{eqnarray}
G_{\mu \nu }+ \lambda g_{\mu \nu } -\frac{1}{2m^2}K_{\mu \nu } = 0,
\label{BHT_eom}
\end{eqnarray}
where
\begin{eqnarray}
K_{\mu \nu }=2\Box R_{\mu \nu } - \frac{1}{2}(\nabla _{\mu }\nabla _{\nu }R + g_{\mu \nu }\Box R) - 8R_{\mu \rho }R^{\rho }_{\ \nu } + \frac{9}{2}RR_{\mu \nu } + g_{\mu \nu }\left ( 3R^{\alpha \beta }R_{\alpha \beta } - \frac{13}{8}R^2 \right ).
\end{eqnarray} 
When a spacetime has a constant curvature as 
$R_{\mu\nu\rho\sigma} = \Lambda(g_{\mu\rho}g_{\nu\sigma}-g_{\mu\sigma}g_{\nu\rho})$, 
$K_{\mu\nu}$ in Eq.~\eqref{BHT_eom} is also simplified as 
$K_{\mu\nu} = -\frac{1}{2}\Lambda ^2g_{\mu\nu}$ \cite{Bergshoeff:2009hq, Oliva:2009ip}, and there can be solutions of constant curvature 
with two different curvature radii
\begin{eqnarray}
\Lambda _{\pm } = 2m(m \pm \sqrt{m^2-\lambda} )
\end{eqnarray}
from Eq.~\eqref{BHT_eom}. 
As mentioned in \cite{Oliva:2009ip}, in the special case with $m^2=\lambda$, the theory has a unique maximally symmetric solution, since the two curvature constants $\Lambda _+$ and $\Lambda _-$ take the same value, $\Lambda _+ = \Lambda _-=2\lambda = 2m^2$. 
For simplicity, we focus on this case in the rest of this paper. 

In this spacial case of the BHT massive gravity, we can obtain the following static circularly symmetric (SCS) black hole solution \cite{Oliva:2009ip}: 
\begin{align}
ds^2 &= -f(r)dt^2+\frac{dr^2}{f(r)}+r^2d\phi ^2, \nonumber \\
f(r) &= -\Lambda r^2 + br - \mu,
\label{SCSsln}
\end{align}
where $\mu $ is related to the mass of the black hole and $b$ is called the gravitational hair parameter \cite{Barnich:2015dvt} which is the origin of the repulsive behavior of the gravity as we will explain later. 
When $b=0$, the solution~\eqref{SCSsln} reduces to the non-rotating BTZ black hole solution. 
As is well known, the BTZ black hole is the unique, nontrivial black hole solution for Einstein gravity in $(2+1)$ dimensions with a negative cosmological constant and $\mu$ can be regarded as the mass parameter in the BTZ black hole solution. 
The black hole mass $M$ is encoded to 
\begin{equation}
M = \frac{\mu+1}{4G},
\end{equation} 
with respect to the AdS spacetime ($b=0$ and $\mu = -1$). 
$\Lambda$ in Eq.~\eqref{SCSsln} works as an effective cosmological constant, while the original cosmological constant $\lambda$ differs twice as much as $\Lambda$. 
Though there is no black hole solution in Einstein gravity when $\Lambda >0$, there exists a black hole solution in the BHT massive gravity  even if $\Lambda >0$ \cite{Oliva:2009ip}. 
In this paper, however, we set $\Lambda < 0$ to compare with the BTZ black hole and consider the case of $b>0$ and $\mu > 0$. 

\begin{figure}[t]
  \begin{center}
\includegraphics[clip, width=15cm]{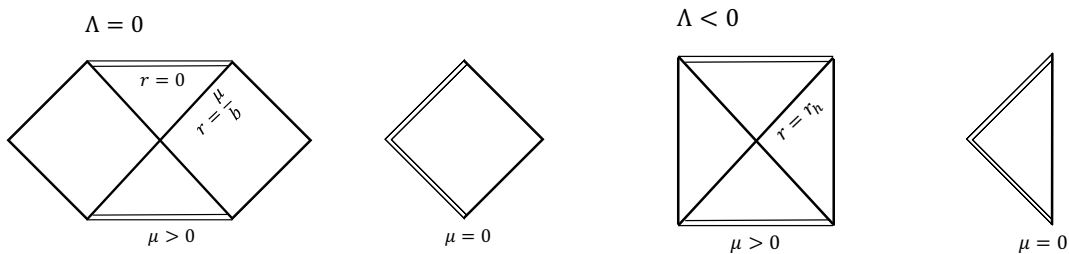}
 \caption{Penrose diagrams for the SCS black hole solution in the BHT massive gravity with various 
    parameters.} 
    \label{penrose}
      \end{center}
\end{figure}

The black hole spacetime represented by the metric~\eqref{SCSsln} has the horizon at 
\begin{eqnarray}
r_{h} = \frac{-b + \sqrt{b^2 -4\Lambda \mu}}{-2\Lambda},
\end{eqnarray}
and its scalar curvature $R$ is calculated as 
\begin{eqnarray}
R = 6\Lambda - \frac{2b}{r}.
\label{scalarcurvature}
\end{eqnarray}
Equation~\eqref{scalarcurvature} means that this spacetime is asymptotically AdS and there is a curvature singularity at $r=0$. 
When $\Lambda =0$, the black hole spacetime represented by the metric~\eqref{SCSsln} is asymptotically locally flat \cite{Oliva:2009ip}. 
Figure \ref{penrose} shows the causal structures of the spacetimes denoted by the metric~\eqref{SCSsln} with some values of parameters. 
The causal structures for the other values of the parameters are discussed in \cite{Barnich:2015dvt}. 
The existence of the gravitational hair $b$ is one of the crucial differences between the BHT massive gravity and Einstein gravity. 
Interestingly, $b$ enables us to find other types of black hole solutions. 
For example, there is a black hole solution with radius $r=\mu /b$ even when $\Lambda =0$ as long as $b$ and $\mu $ are positive \cite{Barnich:2015dvt, Troessaert:2015syk, Alkac:2016xlr}. 
Black hole solutions in the general case, i.e., $m^2 \neq \lambda$ have also been found, e.g., the Lifshitz black hole~\cite{AyonBeato:2009nh} and black holes in the AdS background~\cite{Clement:2009gq,Clement:2009ka}.

\begin{figure}[t]
 \begin{center}
  \includegraphics[width=0.6\textwidth]{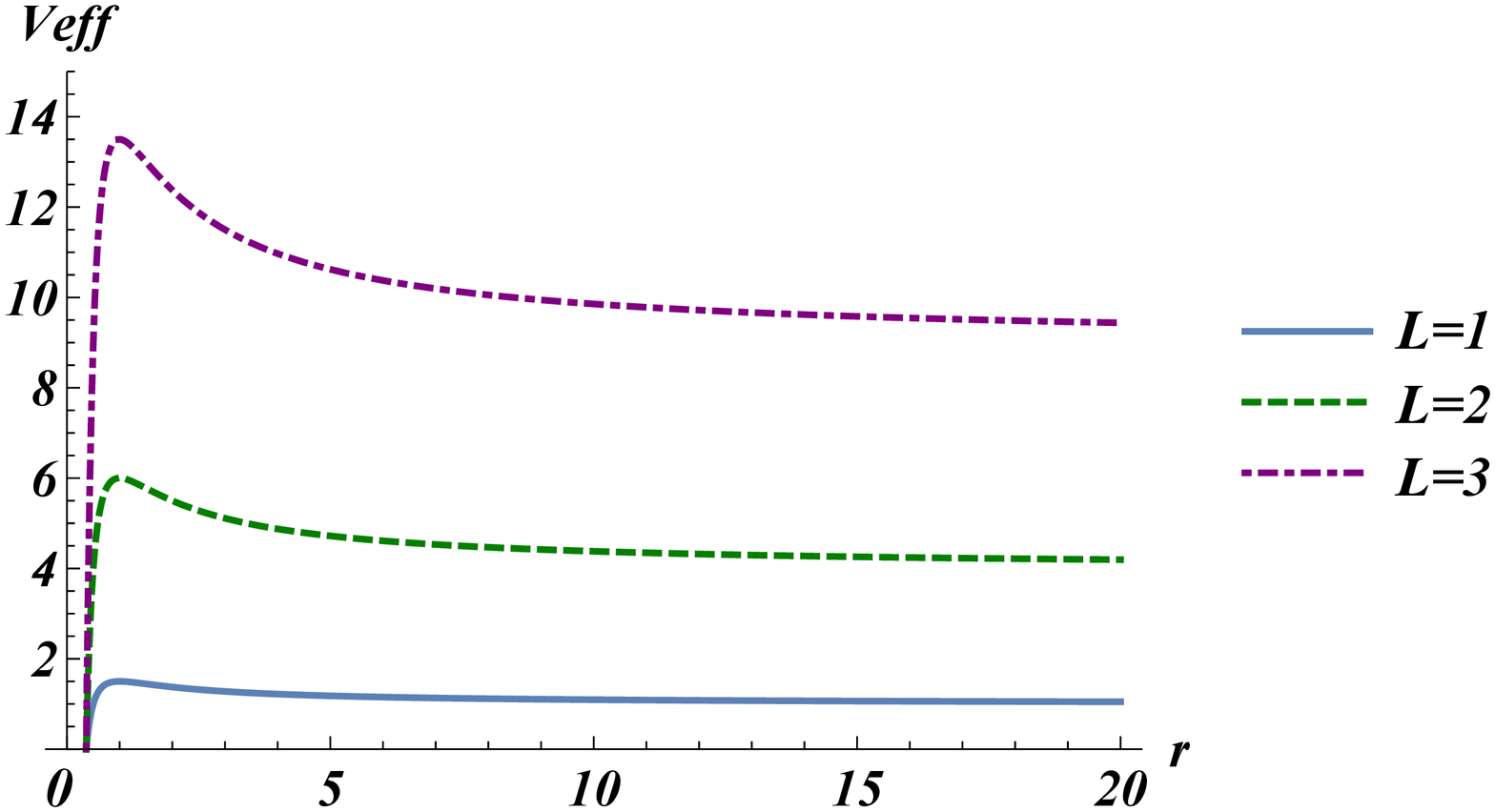}
  \caption{Effective potentials with various angular momenta $L$. }
    \label{Veff}
 \end{center}
\end{figure}

\resection{Null geodesics of the SCS black hole spacetime in the BHT massive gravity}

\ \ \ In this section, we study the null geodesics in the SCS black hole spacetime~\eqref{SCSsln}. 
There are two Killing vectors associated with the spacetime, $(\partial _t)^\mu $ and $(\partial _\phi )^\mu$. 
Thus we can find the constants of motion along the geodesic as follows 
\begin{eqnarray}
E = -g_{\mu t} p^\mu = f(r)\dot t,
\label{energy}
\end{eqnarray}
and 
\begin{eqnarray}
L = g_{\mu \phi }p^\mu = r^2\dot \phi,
\label{amomentum}
\end{eqnarray}
where $p^\mu $ is the four-momentum of a massless particle. 
The dot denotes the derivative with respect to an affine parameter $\lambda$. 
$E$ and $L$ correspond to the energy and the angular momentum of a massless particle, respectively. 
The geodesic equation reduces to an ordinary differential equation by using the null condition $p_\mu p^\mu =0$ as 
\begin{eqnarray}
\dot r^2 = E^2 - f(r)\frac{L^2}{r^2}.
\label{radialEOM}
\end{eqnarray}
Now we define the effective potential $V^2_{\mathrm{eff}}(r)$ by
\begin{eqnarray}
V^2_{\mathrm{eff}}(r) = f(r)\frac{L^2}{r^2}.
\label{effpotential}
\end{eqnarray}   
In Fig. \ref{Veff} we depict the effective potentials for some values of the angular momentum $L$. 
The effective potential has a maximum that corresponds to an unstable circular orbit of a massless particle around the SCS black hole that locates at $r_a=2\mu/b$. 
The maximum value $V_{\mathrm{max}}^2$ of the effective potential increases as $L$ becomes larger.

Combining the geodesic equation~\eqref{radialEOM} and $\dot \phi = L/r^2$, we obtain 
\begin{eqnarray}
\left ( \frac{dr}{d\phi } \right )^2 =\frac{r^4}{L^2}\left ( E^2 - f(r)\frac{L^2}{r^2} \right ) = \frac{E^2}{L^2}r^4 - (-\Lambda r^2 + br - \mu )r^2.
\label{geoeq}
\end{eqnarray}
We can integrate this equation easily.  
The solutions are classified into the following three types by values of the parameters $\bar D$, $\beta $, and $\mu$ (see Fig. \ref{potentialb0}): 
\begin{numcases}
{}
  r_\mathrm{I}(\phi ) = \cfrac{2\mu }{b+2\mu \kappa _\mathrm{I} \sinh (\pm \sqrt{\mu }\phi + \beta )},\ \ \ ({\rm{Type \ I}}: E^2>V_{\mathrm{max}}^2) & \\
  r_{\mathrm{II}}(\phi ) = \cfrac{2\mu }{b + 2\mu \kappa _{\mathrm{II}} \cosh (\pm \sqrt{\mu }\phi + \beta)},\ \ \ ({\rm{Type \ II}}: E^2<V_{\mathrm{max}}^2,\ \ r_0<r_a)& \\
  r_{\mathrm{III}}(\phi ) = \cfrac{2\mu }{b - 2\mu \kappa _{\mathrm{II}} \cosh (\pm \sqrt{\mu }\phi + \beta)},\ \ \ ({\rm{Type \ III}}: E^2<V_{\mathrm{max}}^2,\ \ r_0>r_a) &
\end{numcases}
where $r_0$ is an initial location of a massless particle. 
Also, $\kappa _\mathrm{I} ^2 = (4 \mu/\bar D^2 -b^2)/4 \mu ^2$, $\kappa _{\mathrm{II}} ^2 = (b^2-4\mu/\bar D^2)/4 \mu ^2$ and 
  \begin{align}
    \bar D^2 = \frac{D^2}{1+ D^2 \Lambda}, \ \ \ D = \frac{L}{E},
    \label{impactpara}
  \end{align}
where $D$ is the impact parameter. For the geodesics of Type I, the effective impact parameter $\bar D$ satisfies $4 \mu/\bar D^2 >b^2$ and for the geodesics of Types II and III, $b^2>4\mu/\bar D^2 $. $\beta $ is an integration constant. 
The behaviors of the null geodesics of Types I, II, and III are summarized as follows: 
\begin{figure}[t]
 \begin{center}
  \includegraphics[width=0.5\textwidth]{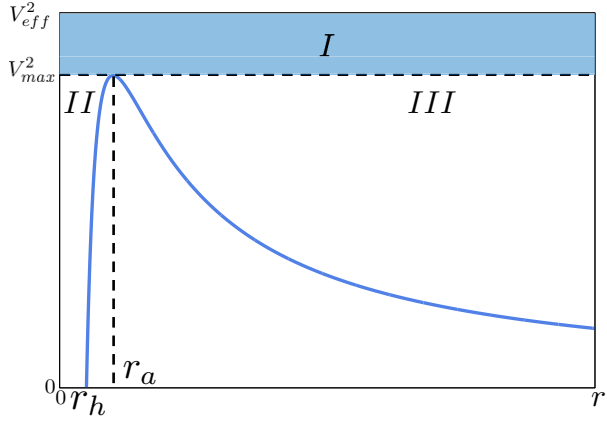}
  \caption{Effective potential for a massless particle in the SCS spacetime. 
  $r_h$ and $r_a$ correspond to the radius of the event horizon and the unstable circular orbit of a massless particle, respectively. }
 \label{potentialb0}
 \end{center}
\end{figure}

\begin{itemize}
\item Type I: the energy of a particle is larger than the maximum value of the effective potential. 
The particle can cross the event horizon from the outside and hits the singularity at the center of the black hole. 
\item Type II: all particles fall to the singularity at the center.
\item Type III: starting from outside the horizon, a particle gets close to the black hole and then moves away from it. 
\end{itemize}
\begin{figure}[t]
  \begin{center}
  \includegraphics[clip, width=13cm]{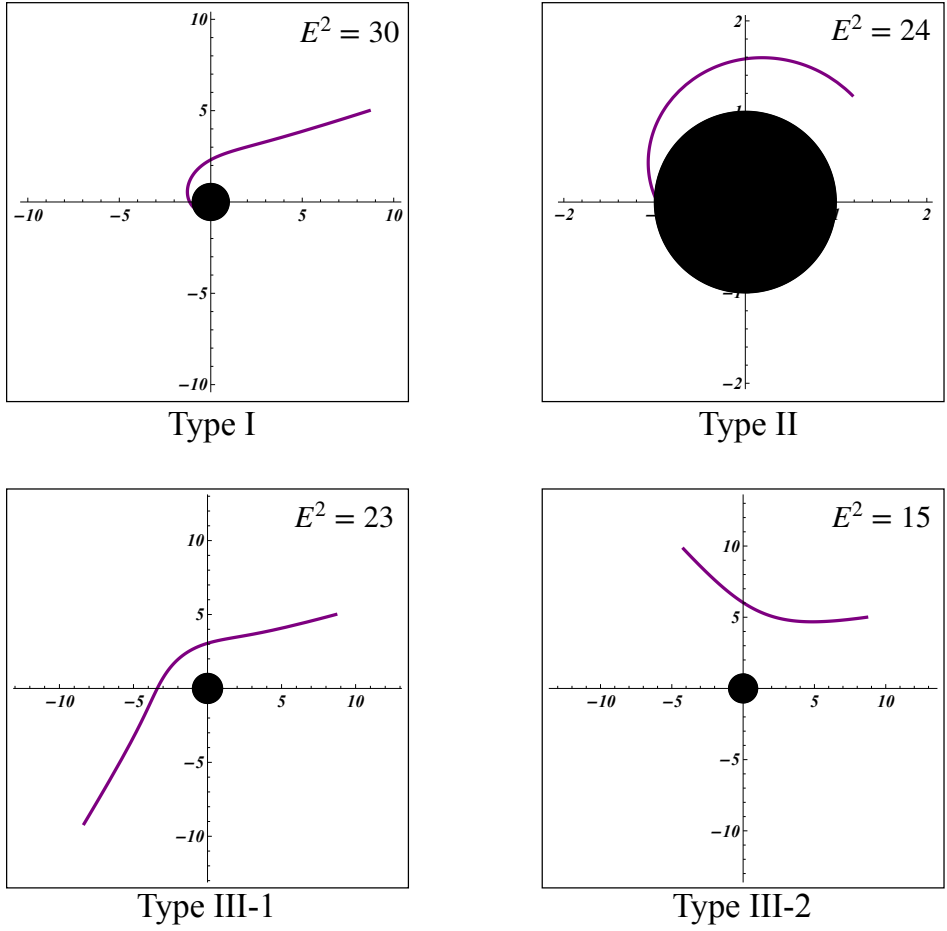}
\caption{Behaviors of the geodesics for various parameters. We fix the parameters as $\Lambda=0,\ b=1,\ \mu=1,\ L=10$.}
    \label{geogeo}
  \end{center}
\end{figure}
We examine the details of the orbits of these solutions by the particles' energy and initial positions as shown in Fig. \ref{geogeo}.  
The null geodesics of Types I and II always fall into the black hole across the horizon. 
In particular, the null geodesics of Type III is worth focusing on. 
The null geodesics of Type III get close to the horizon and can escape to infinity, reflected by the wall of the effective potential. 
However, there are two subclasses of the orbits of this type. 
One is the orbit that goes around the black hole (Type III-1 in Fig. 4) and the other is the orbit that is bounced off from the black hole (Type III-2 in Fig. 4). 
So we can classify all types of geodesics as follows:
\[
  \text{Energy} \left \{ \begin{array}{l}
    \text{lager that the maximum of $V^2_{\mathrm{eff}}$ (I)} \\
    \text{smaller that the maximum of $V^2_{\mathrm{eff}}$}  \left \{ \begin{array}{l} 
    \text{cannot escape to infinity (II)} \\
    \text{can escape to infinity (III) } \left \{ \begin{array}{l}
    \text{go around (III-1)}\\
    \text{be bounced off (III-2)}
    \end{array} \right .
    \end{array} \right .
  \end{array} \right . 
\]
This fact means that the gravity works repulsively on massless particles when the energy takes a value in some range that corresponds to the geodesics of Type III-2. 
We will show this explicitly in next section.
Such behaviors of the geodesics in black hole spacetimes cannot be seen in Einstein gravity where the gravitational force always works attractively. \\

\resection{Deflection angle}
\ \ \ In this section we explicitly calculate the deflection angles of the null geodesics in the SCS black hole spacetime represented by the metric~\eqref{SCSsln} to characterize the repulsive behavior of the gravity. 
We can rewrite the geodesic equation for massless particles (\ref{geoeq}) as
\begin{align}
\left ( \frac{dr}{d\phi } \right )^2 &= \frac{1}{\bar D^2}r^4 - br^3 + \mu r^2, 
\end{align}
where $\bar D $ is defined by Eq.~\eqref{impactpara}. 
As we showed in the previous section, we should classify the null geodesics not only by the energy of a particle 
but also the impact parameter $D$. 
Here, let $D_c$ denote the value of $D$ corresponding to the unstable circular orbit, which is given by $D_c = 2\sqrt{\mu }/\sqrt{b^2 - 4 \mu \Lambda}$. 
The geodesics with $D<D_c$ fall into the black hole, which corresponds to the geodesics of Type I. 
On the other hand, the geodesics with $D>D_c$ are bounced by the wall of the effective potential, which corresponds to the geodesics of Type III.

To characterize the repulsive behavior of the gravity that acts on the geodesics of Type III, we calculate the deflection angle. 
Introducing a new variable $u$ by $u=1/r$, the geodesic equation becomes 
\begin{align}
\left ( \frac{du}{d\phi } \right )^2 = \frac{1}{\bar D^2} - bu + \mu u^2 = F(u).
\label{fu}
\end{align}
We define the deflection angle $\alpha $ in the present coordinate system as 
\begin{align}
\alpha = 2 \int _0 ^{1/R_0} \frac{du}{\sqrt{F(u)}} - \pi,
\label{formalform}
\end{align}
where $R_0$ is the closest approach radius that is given by 
\begin{align}
R_0 = \frac{b\bar D^2 + \bar D\sqrt{b^2 \bar D^2-4\mu }}{2}.
\end{align}
For $F(u)$ in Eq.~\eqref{fu} we obtain the deflection angle as
\begin{align}
\alpha = \frac{1}{\sqrt{\mu }}\log \left | \frac{b\bar D + 2\sqrt{\mu}}{b \bar D - 2\sqrt{\mu }} \right | -\pi.
\label{deflectionangle}
\end{align}
\begin{figure}[t]
    \begin{center}
    \includegraphics[width=0.5\textwidth]{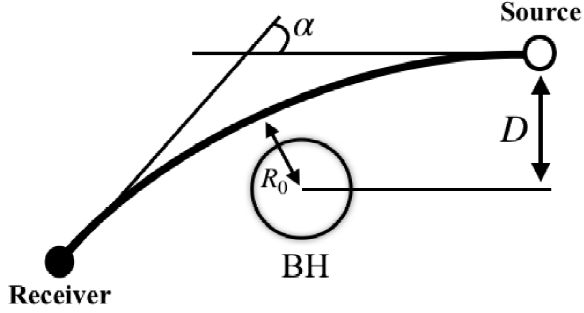}
    \caption{Schematic picture of the deflection angle $\alpha$ and the impact parameter $D$.}
    \label{angle}
    \end{center}
\end{figure} \\
Figure \ref{angle} shows a schematic picture of the deflection angle and the impact parameter. 
In Einstein gravity, the deflection angle is always positive for black holes since the gravity works attractively. 
However, as we have seen in the previous section, the gravity works repulsively for some values of parameters in the SCS black hole spacetime represented by the metric~\eqref{SCSsln}. 
This repulsive behavior of the gravity appears as a negative deflection angle. 
Here, we have to note that in the spacetime~\eqref{SCSsln} there also exists a region where the geodesics receive an attractive force as we can see from the geodesic of Type III-1 in Fig. \ref{geogeo}. 
Actually, we can define the critical value $\bar D_{\alpha =0}$ that corresponds to the border between the repulsive and attractive forces, which is calculated as
\begin{align}
\bar D_{\alpha = 0} = \frac{2\sqrt{\mu }}{b \tanh \frac{\pi \sqrt{\mu }}{2}}.
\label{azero}
\end{align}
For the geodesic with this value, the net deflection angle is equals to zero. 
When $\bar D > \bar D_{\alpha =0}$, the gravity works repulsively and the deflection angle is negative. 
By using Eq.~\eqref{impactpara} we can express $\bar D > \bar D_{\alpha =0}$ in terms of the energy of the particle as
  \begin{align}
    E < L\sqrt{\frac{b^2\tanh ^2 \frac{\pi \sqrt{\mu }}{2}}{4\mu} - \Lambda}.
    \label{ene}
  \end{align}
\footnote{In the case of $b>0$, $\mu >0$ and $\Lambda>0$, the expression under the square root in eq.~\eqref{ene} becomes negative. 
In such a case the maximum value of the effective potential is negative, which means that we cannot define the deflection angle. 
In other cases, e.g., the $b<0$ and $\mu <0$ cases, we cannot define the deflection angle either because there is no unstable circular orbit.  }
Since we consider the case of $b>0$, $\mu >0$, and $\Lambda <0$, the right-hand side is always real. 
Therefore, the orbit with such energy behaves like that of the geodesics of Type III-2 shown in Fig. 4.
In Fig.~\ref{deflectionseries} we summarize these orbits for various values of the impact parameter. 

It is worth mentioning the $b=0$ and $b<0$ cases. 
In both cases, the effective potentials do not have a maximum and they increase monotonically with respect to $r$, which means that there is no potential barrier and all of the ingoing massless particles fall into the black hole. 
This implies that in these cases, we cannot define the deflection angle. 
Therefore, the repulsive behavior of the gravity does not appear in the BTZ black hole background (the $b=0$ case) and it appears only in the the BHT massive gravity with $b>0$.
\begin{figure}[t]
 \begin{center}
  \includegraphics[width=0.6\textwidth]{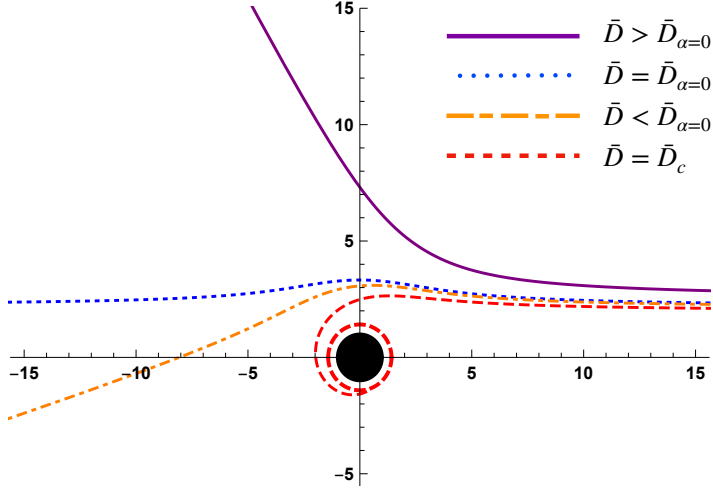}
  \caption{
The goedesics for massless particles for various values of the impact parameter. 
The red dashed line corresponds to the $\bar D_c$ orbit, which takes an unstable circular orbit. 
The orange dot-dashed line is the one for which the gravity works attractively. 
The blue dotted line receives no net gravitational force. 
The purple solid line corresponds to an orbit with a negative deflection angle. 
}
 \label{deflectionseries}
 \end{center}
\end{figure}

\resection{Conclusion and discussion}

\ \ \ In this paper we have studied the null geodesics in the static circularly symmetric black hole in the BHT massive gravity. 
We obtained analytic solutions for the geodesic equation for massless particles and found that the gravity behaves repulsively for the null geodesics with the parameters corresponding to Eq.~\eqref{ene}. 
This repulsive behavior of the gravity cannot be seen by the analysis of the effective potential alone. 
This is because the effective potential tells us only the motion in the radial direction. 

In order to evaluate the repulsive behavior of the gravity in the spacetime represented by Eq.~\eqref{SCSsln}, we have also investigated the deflection angles of the null geodesics. 
We obtained the explicit forms of the deflection angles in terms of the impact parameter $D$, and found that for various values of the impact parameter, the deflection angles can be positive, negative or even zero. 
A negative deflection angle indicates the repulsive behavior of the gravity. 
Note that it is difficult to define the deflection angle in asymptotically non-flat spacetime as many authors are working on (see Ref.~\cite{Ishihara:2016vdc} and references therein). 
The definition that we have used here has been proven to be a scalar quantity at least in the asymptotically flat spacetime~\cite{Ishihara:2016vdc}. 
We would like to investigate whether this definition is also valid in asymptotically non-flat spacetime in future work. 

We have showed that the gravitational hair parameter $b$, which characterizes the BHT massive gravity, is essential for the repulsive behavior of the gravity. 
In fact, when we put $b=0$, the metric~\eqref{SCSsln} reduces to that of the BTZ black hole in Einstein gravity and the repulsive behavior of the gravity does not appear in this case. 

We should note that the sign of the gravitational hair parameter $b$ is important. 
When $b<0$, the effective potential increases monotonically with $r$, so the effective potential does not have a maximum. 
Therefore, all of the ingoing massless particles fall into the black hole, which means that we cannot define the deflection angle for this case. 
The repulsive behavior of the gravity appears only when $b>0$.

Since the BHT massive gravity is a $(2+1)$-dimensional theory, the results that we have obtained in this paper are not directly related to realistic observations.
However, if there is a spacetime whose metric includes a linear term with respect to $r$ such as $br$, it is possible that the gravity could work repulsively for a particle in any dimensions. 
For example, there exist $(3+1)$-dimensional spacetime solutions that have such a linear term in the Weyl conformal gravity theory \cite{Mannheim:1988dj} or in the dRGT massive gravity theory~\cite{Ghosh:2015cva,Chougule:2018cny}. 
The deflection angle for the black hole spacetime in the Weyl conformal gravity was investigated in Refs.~\cite{Ishihara:2016vdc,Edery:1997hu} though the repulsive behavior of the gravity was not noted. 
As in our case, the linear term in the black hole solutions of the Weyl conformal gravity or the de Rahm-Gabadadze-Tolley (dRGT) massive gravity theory makes the deflection angles smaller than those of the Schwarzschild black hole spacetime (see, e.g., Eq.~(12) in Ref.~\cite{Edery:1997hu}). 
The explicit form of the deflection angles that we have obtained in this paper is applicable to strong gravitational lensing as well \cite{Virbhadra:1999nm,Bozza:2001xd,Bozza:2002zj}.  
The possibility of observations of the repulsive behavior of the gravity would be more interesting in $(3+1)$ dimensions; this will be further investigated in a forthcoming paper.

\section*{Acknowledgements}
We would like to thank H. Asada, T. Harada, T. Igata, H. Ishihara, S. Hirano, T. Kobayashi, Y. Koga and N. Tsukamoto for helpful discussions.

\bibliographystyle{JHEP}
\bibliography{refGeodesicBHT}

\end{document}